\documentclass[12pt]{iopart}
\usepackage{latexsym}
\usepackage{epsfig}
\usepackage{amsfonts}
\usepackage{amssymb}
\usepackage{graphicx}

\newcommand{\beq}{\begin{equation}}
\newcommand{\eeq}{\end{equation}}
\newcommand{\bea}{\begin{eqnarray}}
\newcommand{\eea}{\end{eqnarray}}

\def\l{\langle}

\def\f{f}

\def\k{{\bf k}}

\def\curv{\sigma}
\def\f{f_c}

\def\n{{\bf \hat n}}
\def\tf{\tilde f}

\def\l{l}

\begin{document}
\title[Hunting for Isocurvature Modes in the CMB non-Gaussianities]{Hunting for Isocurvature Modes in the Cosmic Microwave Background non-Gaussianities}
\author{David Langlois$^{1,2}$ and Bartjan van Tent$^3$}
\address{$^1$ APC (CNRS-Universit\'e Paris 7), \\
10 rue Alice Domon et L\'eonie Duquet, 75205 Paris Cedex 13, France}
\address{$^2$
 IAP, 98bis Boulevard Arago, 75014 Paris, France}
\address{
$^3$ Laboratoire de Physique Th\'eorique, Universit\'e Paris-Sud 11 et CNRS, \\
B\^atiment 210, 91405 Orsay Cedex, France}

\begin{abstract}
We investigate new shapes 
(in multipole space) of local primordial non-Gaussianities in the CMB. Allowing for a primordial isocurvature mode along with the main adiabatic one, the angular bispectrum is in general a superposition of six distinct shapes: the usual adiabatic 
term, a purely isocurvature component and four additional components that arise from correlations between the  adiabatic and isocurvature modes.  We present a class of early Universe models in which various hierarchies between these six components can be obtained, while satisfying the present upper bound on the isocurvature fraction in the  power spectrum. Remarkably, even with this constraint,   detectable non-Gaussianity could be produced by isocurvature modes. We finally discuss the prospects of detecting these new shapes  with the Planck data, including polarization.

\end{abstract}


\date{\today} 
\maketitle

\section{Introduction}

The very early Universe being a notoriously difficult territory to explore, it is crucial  to extract as much information as possible from the measurement of cosmological perturbations, via the Cosmic Microwave Background (CMB) and large scale structure observations,  as a way to test or constrain early universe scenarios. 
This is why  primordial non-Gaussianity has been the subject of intense study in the last few years (see e.g. \cite{CQG_NG}).

 So far,  WMAP measurements  of the CMB anisotropies~\cite{Komatsu:2010fb}  have set the present limit $f_{\rm
  NL}^{\rm local} =32\pm 21$ (68\,\% CL) [and $-10 < f_{\rm NL}^{\rm
  local} < 74$ (95\,\% CL)] on the parameter $f_{\rm
  NL}^{\rm local} $ that characterizes the amplitude of the simplest type of non-Gaussianity, namely the local shape. Although WMAP data hint at a possible  deviation from Gaussianity, one must wait for the analysis of the  Planck data in order to  confirm or infirm this trend.

A detection of local primordial non-Gaussianity  would have a tremendous impact on our view of the early Universe. This  would indeed rule out all {\it single} field inflation models, which generate only tiny local non-Gaussianity~\cite{Creminelli:2004yq}, and  give strong support to  scenarios with additional scalar fields,  such as another inflaton,  see e.g. \cite{Tzavara:2010ge},  a curvaton~\cite{curvaton} or a modulaton~\cite{modulaton}, which can easily produce  detectable local non-Gaussianity
(this type of non-Gaussianity is also predicted in other scenarios, for instance the ekpyrotic model, see e.g. \cite{Lehners:2010fy}).

Interestingly,  models with multiple scalar fields open up the possibility to generate, in addition to the usual adiabatic fluctuations,  primordial {\it isocurvature} perturbations, corresponding to fluctuations in the relative particle number densities between cosmological fluids (whether isocurvature fluctuations can persist in the post-inflation era depends  on the details of the thermal history of the Universe). 
The amplitude of such isocurvature modes, which could be correlated with the adiabatic one, is now severely constrained by observations of the CMB power spectrum~\cite{Komatsu:2010fb}.

In the present work, we focus on  local  non-Gaussianity generated by these isocurvature fluctuations,
previously studied in 
~\cite{Bartolo:2001cw,Kawasaki:2008sn,Langlois:2008vk,Kawasaki:2008pa,Hikage:2008sk}. 
Improving on these earlier works, we show  that the presence of an isocurvature mode, in addition to the usual adiabatic one, leads  in general to an  angular bispectrum that consists of the superposition of {\it six  elementary 
 components}: the well-known purely adiabatic bispectrum, a purely isocurvature bispectrum, and four other bispectra that arise from the possible  correlations between the adiabatic and isocurvature mode. 
 Because these six bispectra have  {\it different shapes in $l$-space}, their amplitude can  in principle be measured  in the CMB  and  we have
 estimated, via  a Fisher matrix analysis,   what precision on these six parameters could be reached with the Planck data,
 including polarization.

 A natural question is then  whether realistic early Universe models could generate  these new bispectra with detectable amplitude. To answer this question, we consider a simple class of models, which generates all six bispectra with various hierarchies between their relative amplitudes.    
 Remarkably, some of these models  produce detectable non-Gaussianity dominated by the isocurvature mode, while satisfying  the present upper bound on the isocurvature  fraction in the power spectrum. These considerations provide a strong motivation to look for these new shapes of non-Gaussianity in the CMB data.

\section{Angular bispectra}
Let us thus consider  several primordial modes, denoted collectively by $X^I$. We will later focus on the case of two primordial modes:   the usual adiabatic mode, characterized by the total curvature perturbation $\zeta$ (on uniform 
total
energy density hypersurfaces)  and a single CDM (Cold Dark Matter) isocurvature mode $S\equiv3(\zeta_c-\zeta_r)$, with $c$ and $r$ denoting CDM and radiation, respectively. 

The multipole coefficients of the temperature anisotropies ($\Delta T/T=\sum_{lm} a_{lm} Y_{lm}$) are related to the primordial modes $X^I$ 
  via the corresponding  transfer functions $g^I_l(k)$, so that 
\beq
a_{lm}=4\pi (-i)^l \int \frac{d^3\k}{(2\pi)^3} \left(\sum_I X^I(\k) g^I_l(k)\right) Y^*_{lm}(\hat\k).
\eeq
Following \cite{Komatsu:2001rj}, we substitute this  expression into the angular bispectrum
and obtain
\beq
B^{m_1 m_2 m_3}_{l_1\, l_2\,  l_3} \equiv \langle a_{l_1 m_1} a_{l_2 m_2} a_{l_3 m_3}\rangle ={\cal G}^{m_1m_2m_3}_{l_1l_2l_3}b_{l_1l_2l_3}\,,
\eeq
which is the product of  the Gaunt integral
\beq
{\cal G}^{m_1m_2m_3}_{l_1l_2l_3}\equiv \int d^2\n\,  Y_{l_1m_1}(\n)\,  Y_{l_2m_2}(\n)\,  Y_{l_3m_3}(\n)
\eeq
 and of the so-called reduced bispectrum
\begin{eqnarray}
\label{reduced_bispectrum}
b_{l_1l_2 l_3}=\sum_{I,J,K}  \left(\frac{2}{\pi}\right)^3\int \left(\prod_{i=1}^3k_i^2 dk_i\right)  \ g^I_{l_1}(k_1) g^J_{l_2}(k_2) g^K_{l_3}(k_3) 
\cr
\times \, 
B^{IJK}(k_1,k_2, k_3)
\int_0^\infty r^2 dr j_{l_1}(k_1r) j_{l_2}(k_2 r) j_{l_3}(k_3 r)\ 
\end{eqnarray}
that depends on  the bispectra of the primordial $X^I$:
\beq
\langle X^{I}(\k_1) X^J(\k_2) X^{K}(\k_3) \rangle \equiv  (2 \pi)^3 \delta (\Sigma_i \k_i) B^{IJK}(k_1, k_2, k_3)\,.
 \eeq
 The reduced bispectrum (\ref{reduced_bispectrum}) is here
 the sum of several contributions, thus generalizing the purely adiabatic expression given in \cite{Komatsu:2001rj}.

In most inflationary models, 
the ``primordial'' perturbations $X^I$ (defined during the standard radiation era) can be related to the fluctuations of light primordial fields $\phi^a$, generated  at Hubble crossing during inflation, so that one can write, up to second order, 
\beq
\label{X_I}
X^I= N^I_a\,  \delta\phi^a+\frac12 N^{I}_{ab}\,  \delta\phi^a \delta\phi^b + \dots
\eeq
where the $\delta\phi^a$ can usually be treated as  independent quasi-Gaussian fluctuations, i.e. 
$\langle \delta\phi^a (\k) \delta\phi^b (\k')\rangle=
 (2\pi)^3 \, \delta^{ab}P_{\delta\phi}(k) \, \delta(\k + \k')$, with  $P_{\delta\phi}(k)=2\pi^2k^{-3}(H_*/2\pi)^2$, a star denoting Hubble crossing time.
Using Wick's theorem, this implies that the bispectrum $B^{IJK}$ is of the form
\begin{eqnarray}
\label{bispectrum_fields}
 B^{IJK}(k_1, k_2, k_3)&=&
 \lambda^{I, JK}  P_{\delta\phi}(k_2) P_{\delta\phi}(k_3) 
 +\lambda^{J, KI}  P_{\delta\phi}(k_1) P_{\delta\phi}(k_3)
 \cr
 &&
+\lambda^{K, IJ}   P_{\delta\phi}(k_1)P_{\delta\phi}(k_2)\,, 
  \end{eqnarray}
  with
  $\lambda^{I, JK} \equiv \delta^{ac}\delta^{bd}N^I_{ab} N^J_{c} N^K_{d}$ (the summation over scalar field indices is implicit).
Let us  note that the coefficients $\lambda$  are symmetric under the interchange  of the last two indices. 

After substitution of (\ref{bispectrum_fields}) into (\ref{reduced_bispectrum}), the reduced bispectrum can 
finally 
be written as 
\beq
\label{b_l}
b_{l_1l_2 l_3}=   \sum_{I,J,K}\tf_{\rm NL}^{I,JK}b_{l_1l_2 l_3}^{I,JK}
\eeq
where each contribution is of the form
\begin{eqnarray}
\label{b_IJK}
b_{l_1l_2 l_3}^{I,JK}= 3   \int_0^\infty r^2 dr \, \alpha^I_{(l_1}(r)\beta^{J}_{l_2}(r)\beta^{K}_{l_3)}(r),
\end{eqnarray}
using  $(l_1 l_2 l_3)\equiv [l_1l_2l_3+ 5\,  {\rm perms}]/3!$, 
with   
\bea
\label{alpha}
\alpha^I_{\l}(r)&\equiv& \frac{2}{\pi} \int k^2 dk\,   j_\l(kr) \, g^I_{\l}(k)
\\
\label{beta}
\beta^{I}_{\l}(r)&\equiv& \frac{2}{\pi}  \int k^2 dk \,  j_\l(kr) \, g^I_{\l}(k)\,  P_\zeta(k)\,.
\eea
In the $\beta^{I}_{\l}$, we  use  the adiabatic power spectrum:  
$P_\zeta=(\delta^{ab} N_a^\zeta N_b^\zeta) P_{\delta\phi} \equiv {\cal A}\,  P_{\delta\phi}$,  which implies
$\tf_{\rm NL}^{I,JK}= \lambda^{I, JK} /{\cal A}^2$. We also assume that the coefficients $N^I_a$ are weakly time dependent so that the scale dependence of ${\cal A}^2$ can be neglected. 
Note that
 the  {\it elementary} bispectra (\ref{b_IJK}) depend simply on the power spectrum and the transfer functions 
 whereas the main 
 dependence on the early Universe model is embodied by  the $\tf_{\rm NL}^{I,JK}$.

\begin{figure}[h]
\centering
\includegraphics[width=0.8\textwidth, clip=true]{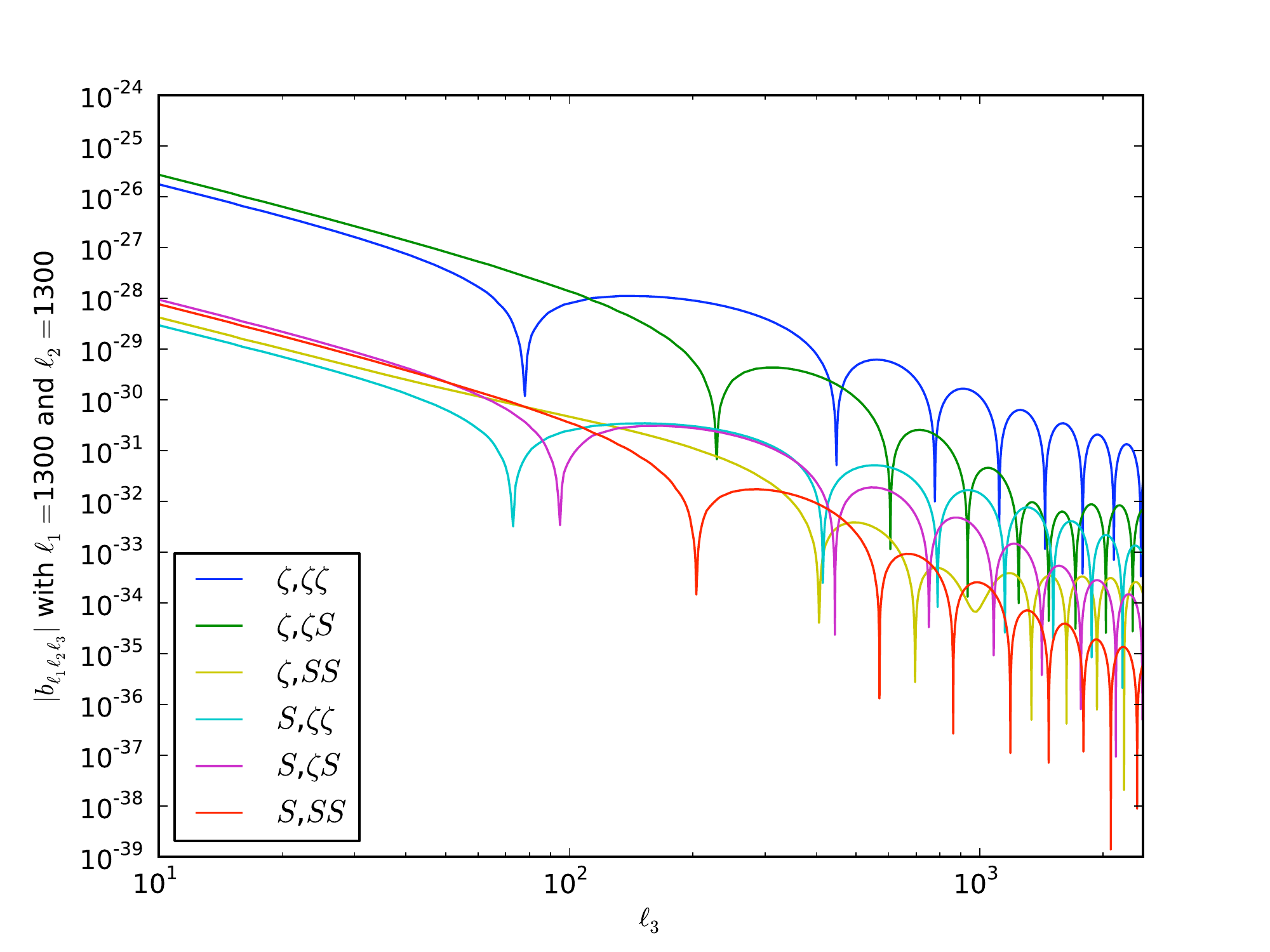}
\caption{Plot of the six elementary bispectra as functions of $l_3$, for $l_1=l_2=1300$.}
\label{fig1}
\end{figure}

If cosmological perturbations depend on  a single scalar field, only the adiabatic mode $\zeta$ exists 
in  (\ref{X_I}),  and (\ref{b_l}) reduces to  the familiar adiabatic bispectrum  with 
$\tf_{\rm NL}^{\zeta,\zeta\zeta}\equiv (6/5) f_{NL}= N^\zeta_{\phi\phi}/(N^{\zeta}_\phi)^2$.
In slow-roll inflation, this coefficient is a combination of the usual slow-roll parameters and is thus too small to be detectable.

However, 
if perturbations are generated by   {\it at least two} light scalar fields, 
an isocurvature mode can later coexist with the adiabatic mode. 
The crucial property, then, is that the {\it adiabatic and isocurvature transfer functions}, which enter into (\ref{alpha}) and (\ref{beta}), are  {\it very different} 
(as illustrated by the respective $C_l=\langle a^*_{lm} a_{lm}\rangle$ plotted in Fig.~\ref{fig}).
 Consequently, the angular bispectrum is now the sum of    {\it six distinct} terms 
 (as illustrated in Fig.\ref{fig1}), with respective weights 
$\tf_{NL}^{\zeta, \zeta\zeta}$, $\tf_{NL}^{\zeta, \zeta S}$, $\tf_{NL}^{\zeta, SS}$, $\tf_{NL}^{S, \zeta\zeta}$, $\tf_{NL}^{S, \zeta S}$, $\tf_{NL}^{S, SS}$.

In the particular case where the adiabatic and isocurvature modes depend on two disjoint subsets of scalar field fluctuations,
the angular bispectrum contains only  a purely adiabatic contribution and  a purely isocurvature one. 
But, in general, the two modes can depend on common scalar field(s), which leads to {\it correlations} between the adiabatic and isocurvature modes. The four mixed contributions to the angular bispectrum must then be taken into account.

\section{Example: a curvaton model}
To  illustrate this  general situation, we  consider a  simple class of models 
based on the presence of a spectator light scalar field   during inflation, dubbed curvaton~\cite{curvaton}. 
This curvaton acquires nearly scale-invariant fluctuations during inflation 
and, later, behaves as a pressureless fluid when it oscillates at the bottom of its potential, before decaying. 

Here, we  allow the curvaton $\curv$  to decay into both radiation and CDM with the respective branching ratios $\gamma_r$ and $\gamma_c$. 
Since, in general, CDM can already be present before the decay, we define the fraction of CDM created by the decay as
$\f\equiv{\gamma_c\,  \Omega_\curv}/({\Omega_c+\gamma_c\Omega_\curv})$, 
where the $\Omega$'s represent the relative abundances just before the decay. 

As shown in \cite{Langlois:2011zz}, the ``primordial'' adiabatic and isocurvature perturbations, i.e. defined 
after the curvaton decay, can be written in the form (\ref{X_I}), with 
\begin{eqnarray}
\label{z_coeffs}
N^\zeta_\curv=\frac{2r}{3\curv_*}, && \quad N^\zeta_{\curv\curv}=\frac{2r}{3\curv_*^2},
\\
 N^S_\curv=\frac{2}{\curv_*}(\f-r), &&
  N^S_{\curv\curv}=\frac{2}{\curv_*^2}\left[ \f(1-2\f)-r \right],
\end{eqnarray}
where  $r\equiv 3\,\gamma_{r } \,  \Omega_\curv/[(4-\Omega_\curv)(1-(1- \gamma_{r}) \Omega_\curv)]$ is assumed to be small, 
  since significant non-Gaussianities arise  only if  $r\ll 1$. 

Let us first discuss linear perturbations. It is  useful to introduce the curvaton contribution to the total adiabatic power spectrum 
$\Xi\equiv (N^{\zeta}_\curv)^2/[(N^{\zeta}_\phi)^2+(N^{\zeta}_\curv)^2]$,  
where $N^\zeta_\phi=H/\dot\phi$ is 
associated with the inflaton fluctuation, and $N^S_\phi=0$. $\Xi$ is  directly related to  the correlation 
${\cal C}\equiv P_{\zeta, S}/\sqrt{P_S P_\zeta}=\sqrt{\Xi}\ {\rm sgn}(\f-r)$. 
The isocurvature-adiabatic ratio,  given by
\beq
\alpha \equiv \frac{P_S}{P_{\zeta}} =\frac{(N^S_\curv)^2}{(N^{\zeta}_\phi)^2+(N^{\zeta}_\curv)^2}= 9\left(1-\frac{\f}{ r}\right)^2 \,\Xi \, ,
\eeq
 is constrained by CMB observations~ \cite{Komatsu:2010fb} to be small (typically $\alpha\lesssim 0.07$)  which requires  either $\f\simeq r$ or $\Xi\ll 1$. 

Let us now turn to non-Gaussianities. Using (\ref{z_coeffs}), one finds $\tf^{\zeta,\zeta\zeta}_{\rm NL}= 3\, \Xi^2/(2r) $. This is the dominant contribution in the 
 regime $\f\simeq r$,  the other components being  suppressed. We thus
concentrate on the more interesting case   $\Xi\ll 1$ 
to discuss the size of the various components in terms of   $\f$ and $r$, considered as free parameters in our phenomenological approach.

In the regime $\f\ll r \ll 1$, the purely adiabatic coefficient is the smallest one. The other ones  are {\it enhanced } by powers of $(-3)$ 
(since $N^S_\sigma/N^\zeta_{\sigma}= N^S_{\sigma\sigma}/N^\zeta_{\sigma\sigma}= -3$):
\beq
\label{f<r}
\tf_{\rm NL}^{I,JK}= (-3)^p \tf^{\zeta,\zeta\zeta}_{\rm NL}\,, \quad  \tf^{\zeta,\zeta\zeta}_{\rm NL}=  \frac{\alpha^2}{54 r}\, ,
\eeq
where  $p$ is the number of ``$S$" in the triplet $\{I,J,K\}$. In particular, the purely isocurvature coefficient is enhanced by a  factor $27$,
 but with the opposite sign: $ \tf^{S,SS}_{\rm NL}=  -\alpha^2/(2 r)$. 
All coefficients can be significant if $r$ is sufficiently smaller than $\alpha^2$.

In the opposite regime $r \ll f_c  \ll 1$, the purely adiabatic coefficient is, once again, the smallest one. All the coefficients  are now positive and enhanced by factors $(3\f/r)^p$, where $p$ is again the number of  ``$S$" indices:
\beq
\label{f>r}
 \tf_{\rm NL}^{I,JK}= \left(\frac{3\f}{r}\right)^p \tf^{\zeta,\zeta\zeta}_{\rm NL}\,,\quad  \tf^{\zeta,\zeta\zeta}_{\rm NL}= \frac{\alpha^2 r^3}{54 \f^4}\,.
\eeq
Note that the enhancement factor is much bigger than in the previous case  (\ref{f<r}). The purely isocurvature coefficient, which dominates, is $ \tf^{S,SS}_{\rm NL}=  \alpha^2/(2 \f)$ 
and can be large if $\f$ is sufficiently small, while the relative size of the other coefficients depends on the ratio $r/\f$.

The above results show   that a  small isocurvature
 fraction in the power spectrum is compatible with a  
 dominantly isocurvature bispectrum 
  detectable by Planck (e.g. $\alpha\simeq 10^{-2}$ and $r\ll f_c\simeq 10^{-8}$ yields $\tf_{\rm NL}^{S,SS}\simeq 5\times 10^3$). 
  Of course, the relations  (\ref{f<r}) or (\ref{f>r}),  are specific to the models considered here and would be a priori different  in other models. 
  It is therefore 
  important
   to try to measure these six coefficients {\it separately}, in order to obtain model-independent constraints from observations.

\section{Observational prospects}
   To estimate these six parameters, which we now denote  $\tf^{(i)}$ , the usual procedure is to minimize 
\beq
\chi^2=\langle (B^{obs}-\sum_i  \tf^{(i)} B^{(i)}), (B^{obs}-\sum_i  \tf^{(i)} B^{(i)})\rangle,
\eeq
where $B$ is  the angle-averaged  bispectrum
$B_{\l_1 \l_2 \l_3} \equiv \sum_{m_i} 
  \left(
\begin{array}{ccc}
\l_1 & \l_2 & \l_3 \cr
m_1 & m_2 & m_3
\end{array}
\right)
 B_{\l_1 \l_2 \l_3}^{m_1 m_2 m_3}$, the matrix denoting the Wigner-3j symbol. The above scalar product is defined  by
$\langle B, B' \rangle\equiv \sum_{l_i}
{B_{l_1l_2l_3}B'_{l_1l_2l_3}}/{\sigma^2_{l_1l_2l_3}}$
with the variance
$\sigma^2_{l_1l_2l_3}\equiv \langle B^2_{l_1l_2l_3}\rangle-\langle B_{l_1l_2l_3} \rangle^2\approx
\Delta_{l_1l_2l_3} C_{l_1}C_{l_2} C_{l_3}$
and  $\Delta_{l_1l_2l_3} =n!$, $n$ being the number of identical indices among $\{l_1, l_2, l_3\}$. For a real experiment the noise power spectrum is added to the $C_l$ in this expression.
The best estimates for the parameters are thus obtained by solving
\beq
\sum_j \langle B^{(i)}, B^{(j)}\rangle \tf^{(j)}=\langle B^{(i)}, B^{obs}\rangle\, ,
\eeq
while the statistical error on the parameters is deduced from the second-order derivatives of $\chi^2$,   which define the Fisher matrix, given in our case by
$F_{ij}\equiv \langle B^{(i)}, B^{(j)}\rangle$. 

We have computed this Fisher matrix by extending the numerical code described in \cite{BucherVanTent} to include isocurvature modes and E-polarization. This treatment takes into account the pure TTT and EEE bispectra, as well as all correlations like TTE and TEE (for more details on the inclusion of polarization see e.g.\ \cite{Yadav:2007rk}). We have taken into account the noise characteristics of the Planck satellite~\cite{Bluebook}, using only the 100, 143, and 217 GHz channels, combined in quadrature. Our computation goes up to $l_\mathrm{max} = 2500$ and uses the WMAP-only 7-year best-fit cosmological parameters~\cite{Komatsu:2010fb}.

\def\text{}
\begin{table}[t]
\begin{center}
\begin{tabular}{|cccccc|}
\hline
$(\zeta, \zeta\zeta)$ & $(\zeta, \zeta S)$ & $(\zeta, S S)$ &  $(S,\zeta\zeta)$  & $(S,\zeta S)$ &  $(S,SS)$\\
\hline
$3.9\times 10^{\text{-2}}$ & $4.5\times 10^{\text{-2}}$ & $2.2\times 10^{\text{-4}}$ &
   $2.1\times 10^{\text{-4}}$ & $6.8\times 10^{\text{-4}}$ & $5.5\times 10^{\text{-4}}$
\\ 
- & $7.2\times 10^{\text{-2}}$ & $5.2\times 10^{\text{-4}}$ &
   $3.6\times 10^{\text{-4}}$ & $1.1\times 10^{\text{-3}}$ & $9.3\times 10^{\text{-4}}$
\\ 
- & - & $3.3\times 10^{\text{-4}}$ &
   $1.7\times 10^{\text{-4}}$ & $3.6\times 10^{\text{-4}}$ & $1.2\times 10^{\text{-4}}$
\\ - & - & - & $1.5\times 10^{\text{-4}}$ & $2.2\times 10^{\text{-4}}$ & $7.9\times 10^{\text{-5}}$
\\- & - & - & - & $5.2\times 10^{\text{-4}}$ & $2.6\times 10^{\text{-4}}$
\\- & - & - & - & - & $2.4\times 10^{\text{-4}}$
\\ \hline
\end{tabular}
\caption{Fisher matrix. Only the upper half coefficients are indicated,  since the matrix is symmetric.}
\label{table}
\end{center}
\end{table}

\begin{figure}[h]
\centering
\includegraphics[width=0.8\textwidth, clip=true]{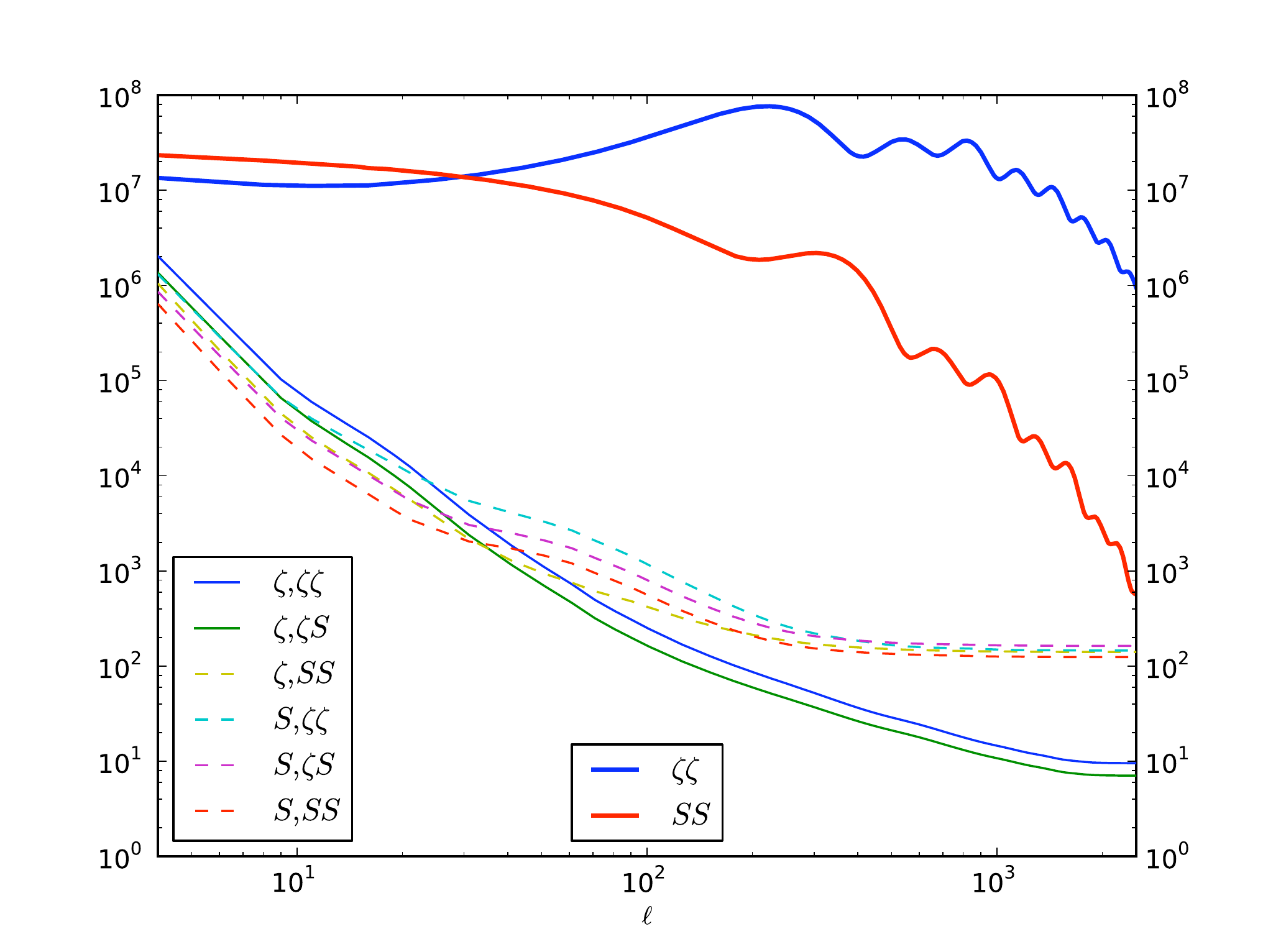}
\caption{The two thick lines are the adiabatic and isocurvature power spectra, more precisely $10^{16} \times l(l+1) C_l/ 2\pi $. The thin lines represent the uncertainties for the six non-Gaussianity coefficients, as functions of the cut-off $\l_{\rm max}$.}
\label{fig}
\end{figure}

From the Fisher matrix given in Table~\ref{table}, 
one finds that the  $68$ \% error on the parameters $\tf^i$ is given by
\beq
\Delta \tf^i=\sqrt{(F^{-1})_{ii}}=\{10, 7, 141,146,163,124\}\,.
\label{errors}
\eeq
The first two uncertainties are much smaller than  the last four. This is due to the severe suppression of  the isocurvature transfer function at high $\l$, which leads to a saturation 
of the signal to noise ratio  for the last  four parameters. By contrast, the large $\l$  behaviour of the first two bispectra  
is governed by the adiabatic transfer function (less severely suppressed), and the error on the first two parameters is thus further reduced at higher $\l$,  
as shown in Fig.~\ref{fig}. 
Including polarization has significantly improved (between $2$ and $6$ times) the precision on the last four parameters.

It is also instructive to  compare (\ref{errors})  with the naive uncertainties  
$\Delta\tf^i= 1/\sqrt{F_{ii}}=\{5, 4, 55, 82,44, 64\}$  
obtained by ignoring the correlations, or, equivalently by assuming that only one parameter  is nonzero. In particular, the contamination of the purely adiabatic signal by the other shapes 
increases
the uncertainty, but only by a factor 2, which is rather moderate
(a two-parameter analysis with  $\tf^{(1)}$ and $\tf^{(6)}$, assuming {\it uncorrelated} adiabatic and isocurvature modes, yields errors almost identical to the single-parameter ones). 
Another consequence of the correlations is that an isocurvature non-Gaussianity could be mistaken for a (much smaller)  adiabatic one by naively using  the purely adiabatic estimator: for instance, a purely isocurvature bispectrum would give a fake  $\tf^{(1)}=  (F_{16}/F_{11})\tf^{(6)}\simeq 10^{-2} \tf^{(6)}$.

\section{Conclusion}
In summary, we have introduced a novel analysis of the CMB data, including polarization, which relies on the decomposition of a generic local adiabatic-isocurvature bispectrum into elementary bispectra. The  corresponding  amplitudes are independent of the details of the early Universe scenario and can be extracted  separately from the data (the correlations between these parameters and  the errors on their measurement are contained in the Fisher matrix which we have computed explicitly for the Planck experiment). The  example of the early Universe scenario discussed here shows that  a detectable CMB bispectrum dominated by isocurvature modes should be considered seriously. 
Looking for isocurvature non-Gaussianity is complementary to the search for an isocurvature component in the power spectrum, which is now standard routine, and it would be highly desirable to conduct systematically both types of analysis for the future CMB data.  

The present study could be extended in several directions. From the observational point of view, 
it would be useful to investigate the contamination 
by foreground emissions or secondary  effects. 
It would also be interesting to see how  the search for isocurvature non-Gaussianities could be improved by combining CMB and large scale structure observations, or even  taking into account  the angular trispectrum~\cite{Kawakami:2009iu,Langlois:2010fe}. From the theoretical point of view, it would be natural to generalize our study to other types of isocurvature modes \cite{LvT_next}. Moreover, it would be worth undertaking a systematic investigation of the 
possible isocurvature non-Gaussianities predicted by various high energy physics scenarios to take advantage of this new observational window.

\ack
The original numerical bispectrum code, which was extended here to include isocurvature modes, was developed by M. Bucher and BvT. We also
acknowledge the use of CAMB.
We would like to thank  R. Crittenden and T. Takahashi for instructive discussions. 
D.L. is partially supported by the ANR grant ÒSTR-COSMOÓ, ANR-09-BLAN-0157.


\section*{References}

\begin{thebibliography}{20}
 
\bibitem{CQG_NG} Focus section on ``Non-linear and non-Gaussian cosmological perturbations'', Class. Quantum Grav., {\bf 27}, 120301 (2010).

\bibitem{Komatsu:2010fb}
  E.~Komatsu {\it et al.} [ WMAP Collaboration ],
  Astrophys.\ J.\ Suppl.\  {\bf 192}, 18 (2011).
  
\bibitem{Creminelli:2004yq}
  P.~Creminelli and M.~Zaldarriaga,
  JCAP {\bf 0410}, 006 (2004)

\bibitem{Tzavara:2010ge}
  E.~Tzavara, B.~van Tent,
 JCAP {\bf 1106}, 026 (2011).

 \bibitem{curvaton}
  A.~D.~Linde and V.~F.~Mukhanov,
  Phys.\ Rev.\  D {\bf 56} 535 (1997) ;
  K.~Enqvist and M.~S.~Sloth,
  Nucl.\ Phys.\  B {\bf 626}, 395 (2002);
  D.~H.~Lyth and D.~Wands,
  Phys.\ Lett.\  B {\bf 524}, 5 (2002);
  T.~Moroi and T.~Takahashi,
  Phys.\ Lett.\  B {\bf 522}, 215 (2001)
  [Erratum-ibid.\  B {\bf 539}, 303 (2002)].
  
  \bibitem{modulaton}
  G.~Dvali, A.~Gruzinov and M.~Zaldarriaga,
  Phys.\ Rev.\  D {\bf 69}, 023505 (2004)
  ;  L.~Kofman,
  arXiv:astro-ph/0303614.
  D.~Langlois and L.~Sorbo,
  JCAP {\bf 0908}, 014 (2009)
  
\bibitem{Lehners:2010fy}
  J.~-L.~Lehners,
  Adv.\ Astron.\  {\bf 2010}, 903907 (2010).
  [arXiv:1001.3125 [hep-th]].  

\bibitem{Bartolo:2001cw}
  N.~Bartolo, S.~Matarrese, A.~Riotto,
  Phys.\ Rev.\  {\bf D65}, 103505 (2002).

\bibitem{Kawasaki:2008sn}
  M.~Kawasaki, K.~Nakayama, T.~Sekiguchi, T.~Suyama and F.~Takahashi,
  JCAP {\bf 0811}, 019 (2008)



\bibitem{Langlois:2008vk}
  D.~Langlois, F.~Vernizzi and D.~Wands,
  JCAP {\bf 0812}, 004 (2008)
    

\bibitem{Kawasaki:2008pa}
  M.~Kawasaki, K.~Nakayama, T.~Sekiguchi, T.~Suyama and F.~Takahashi,
  JCAP {\bf 0901}, 042 (2009)


\bibitem{Hikage:2008sk}
  C.~Hikage, K.~Koyama, T.~Matsubara, T.~Takahashi and M.~Yamaguchi,
  Mon.\ Not.\ Roy.\ Astron.\ Soc.\  {\bf 398}, 2188 (2009)


  
\bibitem{Komatsu:2001rj}
  E.~Komatsu, D.~N.~Spergel,
  Phys.\ Rev.\  {\bf D63}, 063002 (2001).
  
\bibitem{Langlois:2011zz}
  D.~Langlois, A.~Lepidi,
  JCAP {\bf 1101}, 008 (2011).


\bibitem{Bluebook}
Planck Collaboration,
``The Scientific Programme of Planck''  
(2006) [astro-ph/0604069].


\bibitem{BucherVanTent} M.A.~Bucher, B.J.W.~van Tent, and C.S.~Carvalho,
Mon. Not. Roy. Astron. Soc. {\bf 407}, 2193 (2010)

\bibitem{Yadav:2007rk}
 A.~P.~S.~Yadav, E.~Komatsu, B.~D.~Wandelt,
 Astrophys.\ J.\  {\bf 664 } (2007)  680-686.
 [astro-ph/0701921].

\bibitem{Kawakami:2009iu}
  E.~Kawakami, M.~Kawasaki, K.~Nakayama and F.~Takahashi,
  JCAP {\bf 0909}, 002 (2009)
  
\bibitem{Langlois:2010fe}
  D.~Langlois, T.~Takahashi,
  JCAP {\bf 1102}, 020 (2011).
 
 
 \bibitem{LvT_next}
  D.~Langlois and B.~van Tent,
  arXiv:1204.5042 [astro-ph.CO].



 

\end{thebibliography}
\end{document}